\documentclass[titlepage,12pt]{article}
\usepackage{amssymb,amsmath,psfig,epsfig,pslatex}
\usepackage{pstricks,feynmp,psfrag}
\usepackage{graphics,rotating}
\begin{document}
\begin{fmffile}{susypic}
\fmfset{curly_len}{2.5thick} \fmfset{arrow_len}{3thick}
\fmfset{thin}{0.5thick} \fmfset{dot_len}{1mm} \fmfset{wiggly_len}{4thick}
\fmfcmd{%
style_def gluino expr p =
  cdraw (curly p);
  cdraw p
enddef;}

\begin{titlepage}
\noindent
DESY 02-097 \hfill July 2002 \\
PSI-PR-02-06 \hfill
\vspace{0.4cm}
\renewcommand{\thefootnote}{\fnsymbol{footnote}}
\begin{center}
{\LARGE {\bf  Testing coupling relations in \\[0.5cm] SUSY-QCD 
at a Linear Collider{\footnote{Talk presented by A. Brandenburg at the 
10th International Conference on Supersymmetry and Unification of Fundamental
Interactions (SUSY02), June 17-23, 2002, DESY Hamburg.}}}} \\
\vspace{2cm}
{\bf  A. Brandenburg $^{a,}$\footnote
{supported by a Heisenberg fellowship of D.F.G.}, M. Maniatis $^{a}$, 
M. M. Weber $^{b}$}
\par\vspace{1cm}
$^a$  Deutsches Elektronen-Synchrotron DESY, 22603 Hamburg, Germany\\
$^b$  Paul Scherrer Insitut PSI, CH-5232 Villigen, Switzerland
\par\vspace{2cm}
{\bf Abstract:}\\
\parbox[t]{\textwidth}
{Supersymmetry predicts that gauge couplings are
equal to the corresponding gaugino-sfermion-fermion Yukawa couplings.
This prediction can be tested for the QCD sector of the MSSM by studying
the processes $e^+e^-\to$ squark+antisquark+gluon 
and $e^+e^- \to$ squark+antiquark+gluino at a future linear collider.
We present results for these
processes at next-to-leading order in $\alpha_s$ in the framework of the MSSM. 
We find sizable SUSY-QCD
corrections. The renormalization scale
dependence is significantly reduced
at next-to-leading order.}
\end{center}
\vspace{2cm}
\end{titlepage}
\renewcommand{\thefootnote}{\arabic{footnote}}
\setcounter{footnote}{0}
\section{Introduction}
Softly broken supersymmetry predicts that each known particle has a 
superpartner with equal gauge quantum numbers and spin different by
$1/2$. Further, the gauge couplings have to be equal to the corresponding
gaugino-sfermion-fermion Yukawa couplings. These coupling relations 
are vital for the cancellation
of quadratic divergencies. Therefore, in order to establish 
supersymmetry experimentally, 
one not only has to find new particles and measure their quantum numbers, but
also to verify the SUSY coupling relations. For the QCD sector of the MSSM,
the coupling relation is depicted in Fig.~1. Analogous relations hold for the
electroweak sector \cite{freitas}.
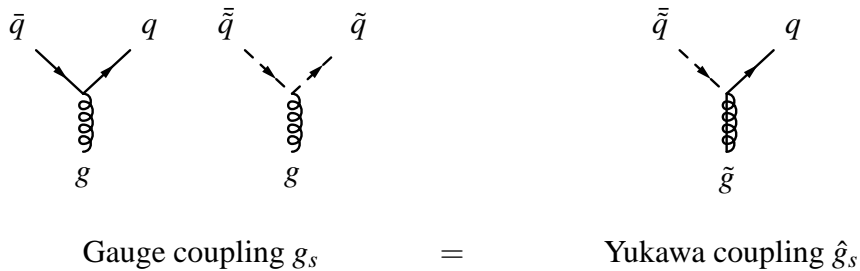
\begin{figure}[b]
\vspace{0.5cm}
\unitlength0.5cm
\begin{center}
\begin{tabular}{ccc}
\fmfframe(0,0)(0,0){
\begin{fmfgraph*}(2.5,3.0)
  \fmfbottomn{i}{1} \fmftopn{o}{2}
  \fmf{gluon,tension=1.5}{i1,v1}
  \fmf{fermion}{o1,v1,o2}
\fmflabel{$\Bar{q}$}{o1}
\fmflabel{$q$}{o2}
\fmflabel{$g$}{i1}
\end{fmfgraph*}}

&
\hspace{1cm}
\fmfframe(0,0)(0,0){
\begin{fmfgraph*}(2.5,3)
  \fmfbottomn{i}{1} \fmftopn{o}{2}
  \fmf{gluon,tension=1.5}{i1,v1}
  \fmf{scalar}{o1,v1,o2}
\fmflabel{$\Bar{\Tilde{q}}$}{o1}
\fmflabel{$\Tilde{q}$}{o2}
\fmflabel{$g$}{i1}
\end{fmfgraph*}}

&
\hspace{4cm}
\fmfframe(0,0)(0,0){
\begin{fmfgraph*}(2.5,3)
  \fmfbottomn{i}{1} \fmftopn{o}{2}
  \fmf{gluino,tension=1.5}{i1,v1}
  \fmf{scalar}{o1,v1}
  \fmf{fermion}{v1,o2}
\fmflabel{$\Bar{\Tilde{q}}$}{o1}
\fmflabel{$q$}{o2}
\fmflabel{$\Tilde{g}$}{i1}
\end{fmfgraph*}}
\end{tabular} \\[1cm]
 \hspace{1.75cm}Gauge coupling $g_s$ \hspace{1.5cm}$=$\hspace{1.75cm}  
Yukawa coupling $\hat{g}_s$  
 \\[0.25cm]
\caption{Coupling relation for the QCD sector of the MSSM.}
\label{couplingrel}
\end{center}
\end{figure}
\par
The SU(3) coupling relation can be verified at $e^+e^-$ colliders by 
measuring the cross sections for the processes
\begin{eqnarray}
 && e^+e^-\to q\Bar{q} g \hspace{2cm} (a),\nonumber\\
 && e^+e^-\to \Tilde{q}\Bar{\Tilde{q}} g \hspace{2cm} (b),\\
 && e^+e^-\to q\Bar{\Tilde{q}} \Tilde{g}, \Bar{q}\Tilde{q} \Tilde{g} 
\hspace{1.2cm} (c),\nonumber
\label{reac}
\end{eqnarray} 
and compare them to precise theoretical predictions, i.e. to a calculation
at next-to-leading order (NLO) in $\alpha_s$. The standard NLO QCD 
corrections to the 3-jet production process  (\ref{reac}a) 
are well known \cite{3jet}. 
Here we will also present the virtual SUSY-QCD corrections to 
order $\alpha_s^2$ for this 
process. Cross sections for processes (\ref{reac}b) and (\ref{reac}c) 
have been computed to leading order in \cite{Schiller}. We have recently
performed a full NLO calculation for these processes. In this talk, we will
discuss our main  results. A detailed account of our work will be
given elsewhere \cite{BMWZ02}.
\section{Squark and gluino production at leading order}
We first discuss the leading order cross sections
for processes (\ref{reac}b) and (\ref{reac}c). 
We allow only light quarks $q=u,d,s,c,b$ 
(and antiquarks)  in the final state of process 
(\ref{reac}c) and neglect their masses.
We further exclude scalar top quarks as final state particles. The mixing
between the chiral components $\Tilde{q}_L$ and $\Tilde{q}_R$  is 
neglected and all five squark flavours are assumed to be mass degenerate.
The cross section for (\ref{reac}b) diverges as the gluon energy goes to zero.
We define an infrared finite cross section as follows:
\begin{eqnarray}
\sigma(E_{\rm cut})\equiv \sum_{q=u,d,s,c,b}\ \sum_{h=L,R}
\sigma(e^+e^-\to \Tilde{q}_h\Bar{\Tilde{q}}_hX; \ E_X>E_{\rm cut}),
\label{sigmacut}
\end{eqnarray}
where $X=g$ at leading order. The minimal energy $E_{\rm cut}$ can be chosen
by the experimentalist. We will take $E_{\rm cut}= 50$ GeV for most of 
the numerical results below.
For process (\ref{reac}c), the total cross section is defined by
\begin{eqnarray}
\sigma_{\rm tot}\equiv \sum_{q=u,d,s,c,b}\ \sum_{h=L,R}
 \Big\{\sigma(e^+e^-\to \Tilde{q}_h\Bar{q}\Tilde{g})+
\sigma(e^+e^-\to q\Bar{\Tilde{q}}_h\Tilde{g})\Big\}.
\label{sigmatot}
\end{eqnarray}
Figs.~2a and 2b show the leading order cross sections $\sigma(E_{\rm cut})$
and $\sigma_{\rm tot}$ defined above. We also plot for comparison 
the LO cross section
for $e^+e^-\to \Tilde{q}\Bar{\Tilde{q}}$. We distinguish between the cases 
$m_{\Tilde{q}}<m_{\Tilde{g}}$ (Fig.~2a) and $m_{\Tilde{q}}>m_{\Tilde{g}}$ 
(Fig.~2b).
One sees that even for rather large squark and gluino masses, the cross 
section $\sigma(E_{\rm cut})$ defined in (\ref{sigmacut})  
reaches values of several tens of femtobarn,
while  the cross section $\sigma_{\rm tot}$ defined in (\ref{sigmatot}) is 
about one order of magnitude smaller. 
 \begin{figure}[h]
\unitlength1.0cm
\begin{center}
\vspace{-2.5cm}
\begin{picture}(8.75,8.75)
\put(-4.,0){\psfig{figure=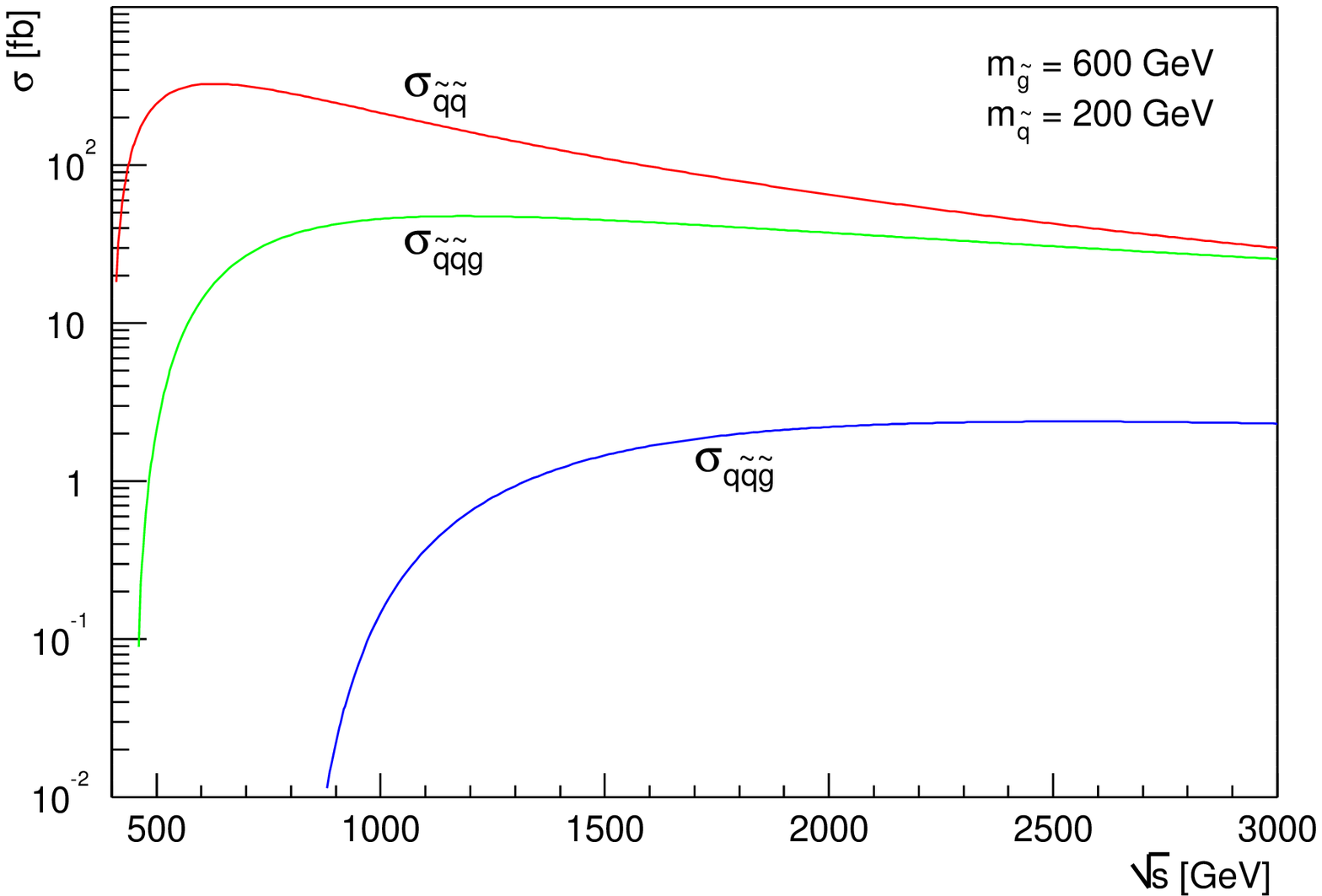,width=8.75cm}}
\put(4.25,0){\psfig{figure=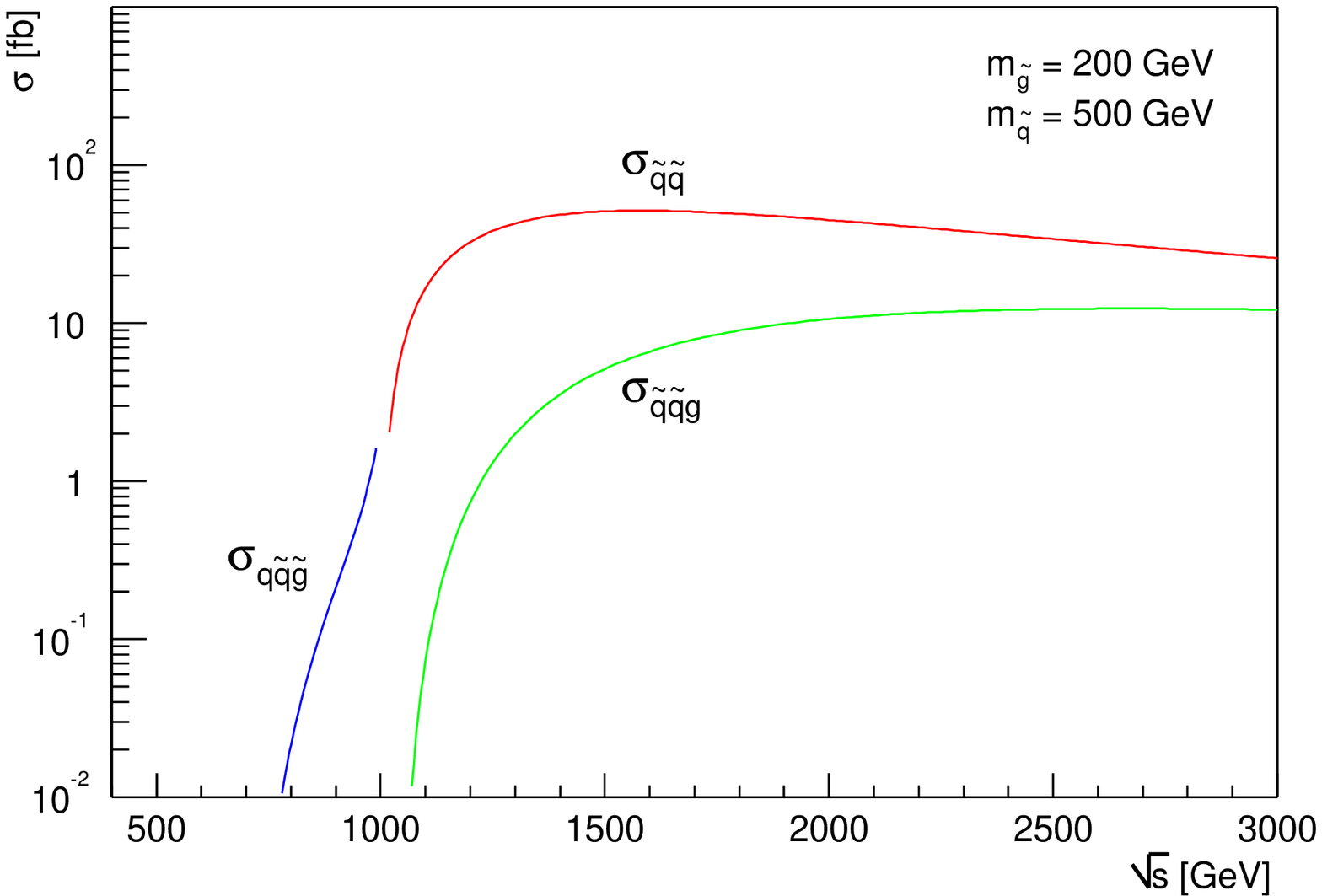,width=8.75cm}}
\put(3,1.25){(a)}
\put(11.25,1.25){(b)}
\end{picture}
\caption{Leading order cross sections for squark and gluino production
in $e^+e^-$ collisions for $m_{\tilde{g}}>m_{\tilde{q}}$ (a) and
$m_{\tilde{g}}>m_{\tilde{q}}$ (b).}
\label{fig:tree}
\end{center}
\end{figure}
\section{SUSY-QCD corrections to  $e^+e^-\to \tilde{q}\bar{\tilde{q}} g$
and  $e^+e^-\to q\bar{\tilde{q}} \tilde{g}/\bar{q}\tilde{q}\tilde{g}$}
The calculation of the NLO SUSY-QCD corrections to processes 
(\ref{reac}b) and (\ref{reac}c) requires the evaluation of both virtual
and real corrections. Possible real corrections at order $\alpha_s^2$ 
to process (\ref{reac}b) are contributions from the processes 
$e^+e^-\to \Tilde{q}\Bar{\Tilde{q}}gg$ and 
$e^+e^-\to \Tilde{q}\Bar{\Tilde{q}}q\Bar{q}$, while the real corrections 
to process  (\ref{reac}c) consist of real gluon emission,
$e^+e^-\to q\Bar{\Tilde{q}}\Tilde{g}g$ . 
Sample Feynman diagrams for both virtual and real
corrections are shown in Figs.~3a-3d.
 \begin{figure}[h]
\vspace{0.5cm}
\unitlength0.5cm
\begin{center}
 \begin{tabular}{cccc}
\fmfframe(0,0)(0,0){
\begin{fmfgraph*}(5,4)
  \fmfleftn{i}{1} \fmfrightn{o}{3}
  \fmf{photon,tension=4,label=$\gamma$,,$Z^0$,label.sid=right}{i1,v1}
  \fmf{phantom,tension=1}{v1,v2}
  \fmf{phantom}{v2,o1}
  \fmf{phantom,tension=1}{v1,v4}
  \fmf{phantom}{v4,o3}
  \fmf{phantom,tension=0.6}{v1,v3}
  \fmf{phantom}{v3,o2}
  \fmffreeze
  \fmf{fermion}{v4,v1,v2}
  \fmf{gluino}{v2,v3,v4}
  \fmf{scalar}{v2,o1}  
  \fmf{scalar}{o3,v4}
  \fmf{gluon}{v3,o2}
 \fmflabel{$g$}{o2}
\fmflabel{$\Tilde{q}$}{o1}
\fmflabel{${\Bar{\Tilde{q}}}$}{o3}
\end{fmfgraph*}}
&
\hspace{0.66cm}
\fmfframe(0,0)(0.66,0){
\begin{fmfgraph*}(5,4)
  \fmfleftn{i}{1} \fmfrightn{o}{4}
  \fmf{photon,tension=3,label=$\gamma$,,$Z^0$,label.sid=right}{i1,v1}
  \fmf{scalar}{o1,v1}
  \fmf{scalar}{v1,o4}
  \fmf{gluon}{v1,v2,o3}  
  \fmflabel{$g$}{o3}
  \fmf{gluon}{v2,o2}  
  \fmflabel{$g$}{o2}
\fmflabel{$\Tilde{q}$}{o4}
\fmflabel{$\Bar{\Tilde{q}}$}{o1}
\end{fmfgraph*}}
&
\hspace{1.33cm}
\fmfframe(0,0)(1.33,0){
\begin{fmfgraph*}(5,4)
  \fmfleftn{i}{1} \fmfrightn{o}{3}
  \fmf{photon,tension=4,label=$\gamma$,,$Z^0$,label.sid=right}{i1,v1}
  \fmf{phantom,tension=1}{v1,v2}
  \fmf{phantom}{v2,o1}
  \fmf{phantom,tension=1}{v1,v4}
  \fmf{phantom}{v4,o3}
  \fmf{phantom,tension=.6}{v1,v3}
  \fmf{phantom}{v3,o2}
  \fmffreeze
  \fmf{scalar}{o1,v2,v1}
 \fmf{scalar}{v1,v4}
  \fmf{quark}{v4,v3,o2}
  \fmf{gluon}{v2,v3}
  \fmf{gluino}{v4,o3}
  \fmflabel{$q$}{o2}
\fmflabel{$\Tilde{g}$}{o3}
\fmflabel{$\Bar{\Tilde{q}}$}{o1}
\end{fmfgraph*}}
&
\hspace{0.66cm}
\fmfframe(0,0)(0.66,0){
\begin{fmfgraph*}(5,4)
  \fmfleftn{i}{1} \fmfrightn{o}{4}
  \fmf{photon,tension=3,label=$\gamma$,,$Z^0$,label.sid=right}{i1,v1}
  \fmf{phantom}{o1,v1}
  \fmf{phantom}{v1,o4}
  \fmffreeze
  \fmf{scalar}{o1,v3,v1}
  \fmf{scalar}{v1,v2}
  \fmf{quark}{v2,o4}
\fmffreeze
  \fmf{gluino}{v2,o3}  
    \fmf{gluon}{v3,o2}
\fmflabel{$\Tilde{g}$}{o3}
\fmflabel{$g$}{o2}
\fmflabel{$q$}{o4}
\fmflabel{$\Bar{\Tilde{q}}$}{o1}
\end{fmfgraph*}}
\\[0.5cm] (a) & (b) &(c) &(d)
\end{tabular}
\\[0.5cm]
\caption{Sample Feynman diagrams for virtual and real corrections
to $e^+e^-\to \tilde{q}\bar{\tilde{q}}g$ (Figs.~ (a) and (b)),
and to  $e^+e^-\to q\bar{\tilde{q}}\tilde{g}$ (Figs.~ (c) and (d)). 
The trivial leptonic part is not shown.}
\label{sample}
\end{center}
\end{figure}
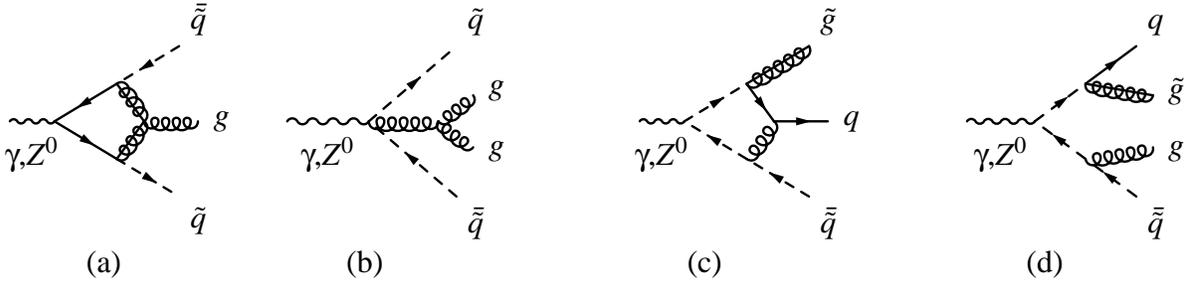
\par
In the calculation of the virtual corrections one encounters both 
ultraviolet and soft and collinear divergencies. For convenience, 
we regularize them by using conventional dimensional regularization (DREG). 
As is well known, DREG violates supersymmetry. In particular, SUSY-invariant
physical quantities involving the gauge and Yukawa couplings $g_s$ and 
$\hat{g}_s$ deviate by a finite amount in the exact SUSY limit.
A SUSY-restoring counterterm can be derived from the Slavnov-Taylor 
identities \cite{hollik01}. 
This amounts to the following replacement of the Yukawa
coupling $\hat{g}_s$ in the gluino-quark-squark vertex \cite{martin}:
\begin{eqnarray}
\hat{g}_s=g_s\left[1+\frac{\alpha_s}{3\pi}\right].
\label{shift}
\end{eqnarray}  
The ultraviolet divergencies are removed by renormalization of the
masses and the coupling. For the mass renormalization we use the on-shell 
scheme, while the strong coupling is renormalized in the $\overline{\rm MS}$ 
scheme. After renormalization the virtual corrections still contain 
soft and collinear singularities 
which manifest themselves in double 
and single poles in $\epsilon=(d-4)/2$. 
They
are cancelled by singular contributions from the four parton final states
which are due to soft gluons and collinear massless partons. These 
contributions have to be computed analytically in $d$ dimensions to 
perform the cancellation. For process (\ref{reac}b) we used the so-called
dipole subtraction method \cite{catani1}, which has been recently generalized  
to include massive partons \cite{catani2}. For process (\ref{reac}c) we used
a variant of the phase space slicing method \cite{giele}. 
\begin{figure}[h]
\begin{center}
\vspace{-0.5cm}
\psfrag{sqrts}{$\sqrt{s}$ [GeV]}
\psfrag{cross section}{$\sigma(E_{\rm cut})$ [fb]}
\psfrag{process}{}
\vspace{-0.5cm}
\epsfig{file=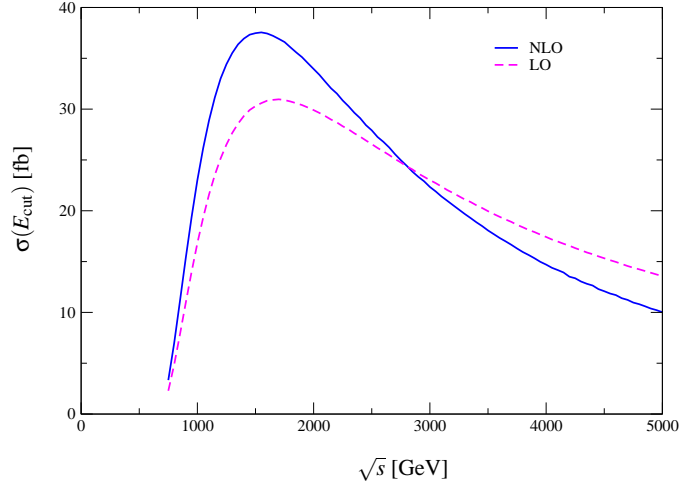,angle=270,width=10cm}
\vspace{-0.25cm}
\caption{Cross section for $e^+e^-\to \tilde{q}\bar{\tilde{q}}X$ 
at leading order (dashed line) and including the SUSY-QCD corrections
(full line) for $m_{\tilde{g}}=400$ GeV,
$m_{\tilde{q}}=300$ GeV, $E_{\rm cut}=50$ GeV and $\mu=1$ TeV.}
\label{markos1}
\end{center}
\end{figure}
 \begin{figure}[h!]
\unitlength1.0cm
\begin{center}
\vspace{-3.5cm}
\begin{picture}(8.75,8.75)
\psfrag{muq}{$\mathsf{\frac{\mu}{\mu_0}}$}
\psfrag{cross section}{$\sigma(E_{\rm cut})$ [fb]}
\psfrag{Ecut10}{$E_{\rm  cut}=10$ GeV}
\psfrag{Ecut100}{$E_{\rm cut}=100$ GeV}
\put(-4,6){\epsfig{figure=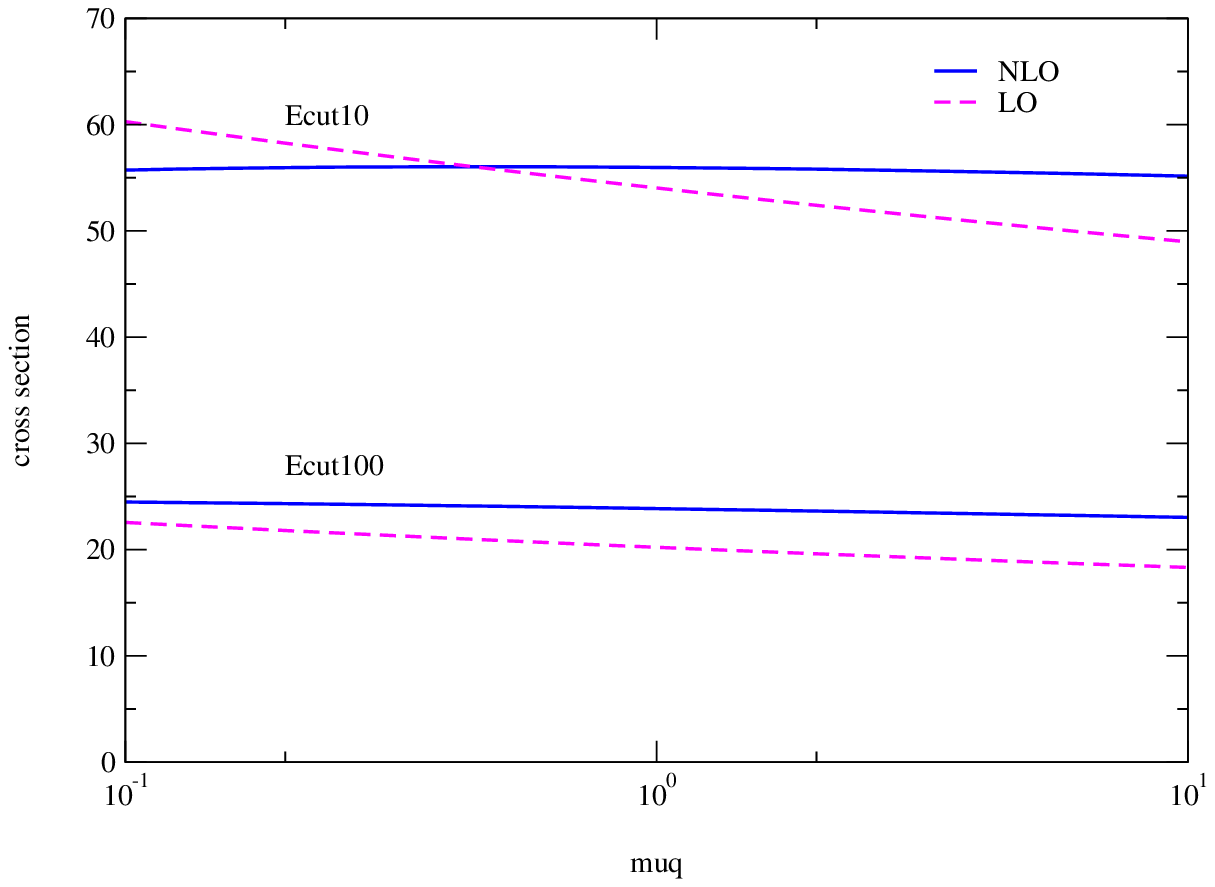,angle=270,width=8.75cm}}
\put(-2.5,0.8){(a)}
\psfrag{cutE}{$E_{\rm cut}$ [GeV]}
\psfrag{sqrts=2000}{}
\put(4.25,6){\epsfig{figure=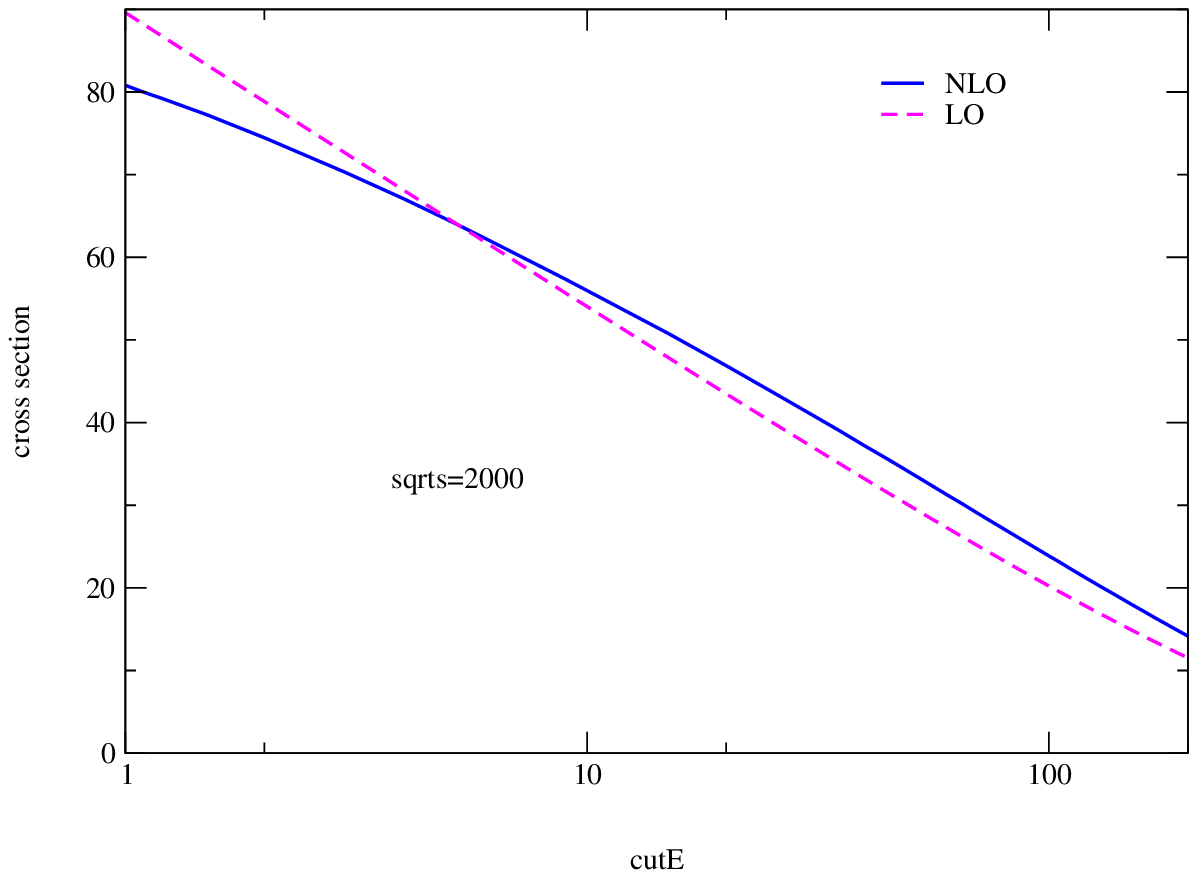,,angle=270,width=8.75cm}}
\put(5.75,0.8){(b)}
\end{picture}
\vspace{0.75cm}
\caption{In (a), the dependence of $\sigma(E_{\rm cut})$ on the 
renormalization scale is shown, where the reference scale is $\mu_0=1$ TeV. 
In (b), the dependence on $E_{\rm cut}$ 
is depicted for $\mu=1$ TeV. The c.m. energy was fixed
to be $\sqrt{s}=2$ TeV, and masses were chosen as in Fig.~4.}
\label{markos2}
\end{center}
\end{figure}
\par
We now discuss our numerical results for the SUSY-QCD corrections, first
for the process $e^+e^-\to \Tilde{q}\Bar{\Tilde{q}}g$.
Fig.~4 shows the cross section $\sigma(E_{\rm cut})$ defined in
Eq.~(\ref{sigmacut}), where at NLO the unspecified part of the final state
$X$ can be a single gluon, two gluons or a massless $q\Bar{q}$ pair.
In all cases we require $E_X>E_{\rm cut}$ and take in Fig.~4
$E_{\rm cut}=50$ GeV. For the masses we choose $m_{\Tilde{g}}=400$ GeV and
$m_{\Tilde{q}}=300$ GeV. 
The renormalization scale is set to $\mu=1$ TeV.
Fig.~4 shows that the SUSY-QCD corrections enhance the cross section in the 
peak region by about 20\%. 
\par
In Fig.~5a we show the dependence of $\sigma(E_{\rm cut})$ on the 
renormalization scale $\mu$ for fixed $\sqrt{s}=2$ TeV and two different
values for $E_{\rm cut}$. The $\mu$-dependence is significantly reduced 
by the inclusion of the order $\alpha_s^2$ corrections, thus leading 
to a more reliable theoretical prediction of the cross section.
Fig.~5b shows the dependence of $\sigma(E_{\rm cut})$ on the cut parameter.
We emphasize again that $E_{\rm cut}$ is a physical parameter that can 
be chosen by the experimentalist. 
\par
The inclusive cross section at LO and NLO for  
$e^+e^-\to q\Bar{\Tilde{q}}\Tilde{g}/\Bar{q}\Tilde{q}\Tilde{g}$ 
defined in (\ref{sigmatot}) 
is shown in Fig.~6a and 6b for the cases $m_{\Tilde{q}}<m_{\Tilde{g}}$
and  $m_{\Tilde{q}}>m_{\Tilde{g}}$, respectively. 
The cross section $\sigma_{\rm tot}$ reaches values up to 
5 fb for the case $m_{\Tilde{q}}=300$ GeV and $m_{\Tilde{g}}=400$ GeV. 
The SUSY-QCD corrections enhance the cross section in the peak region
by about 25\%. For $m_{\Tilde{q}}>m_{\Tilde{g}}$, the cross section is below
1 fb, and this case is probably of no phenomenological interest. 
The NLO corrections reach values up to 80\%{\footnote{For $m_{\Tilde{q}}>m_{\Tilde{g}}$  the process 
$e^+e^-\to q\Bar{\Tilde{q}}\Tilde{g}/\Bar{q}\Tilde{q}\Tilde{g}$ 
is only relevant for c.m.
energies between $m_{\Tilde{q}}+m_{\Tilde{g}}$ and $2m_{\Tilde{q}}$, 
since for $\sqrt{s}>2m_{\Tilde{q}}$  on-shell production of a 
squark-antisquark pair becomes possible.}}.
 \begin{figure}[h]
\unitlength1.0cm
\begin{center}
\vspace{-3.5cm}
\begin{picture}(8.75,8.75)
\put(-4.,0){\psfig{figure=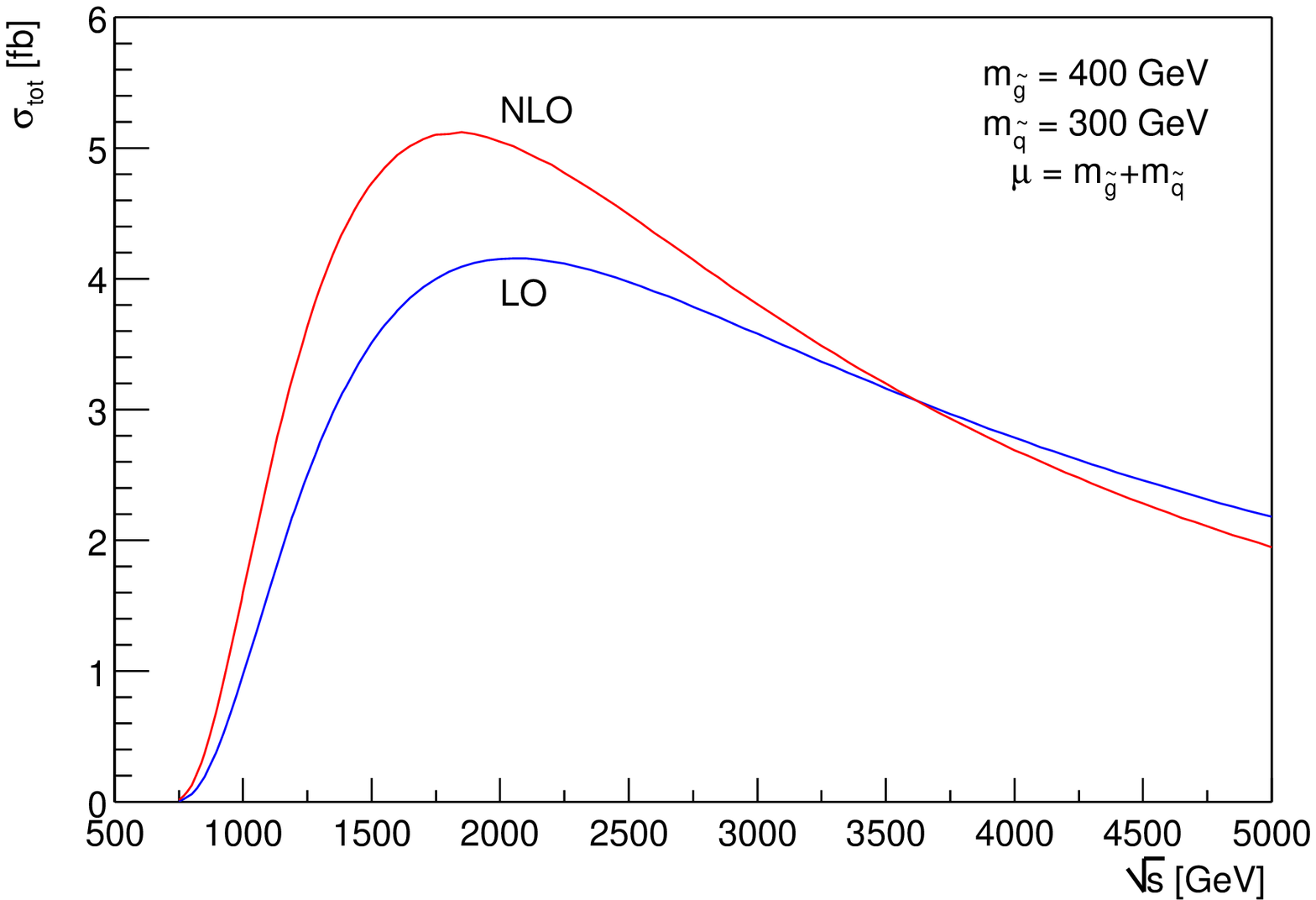,width=8.75cm}}
\put(4.25,0){\psfig{figure=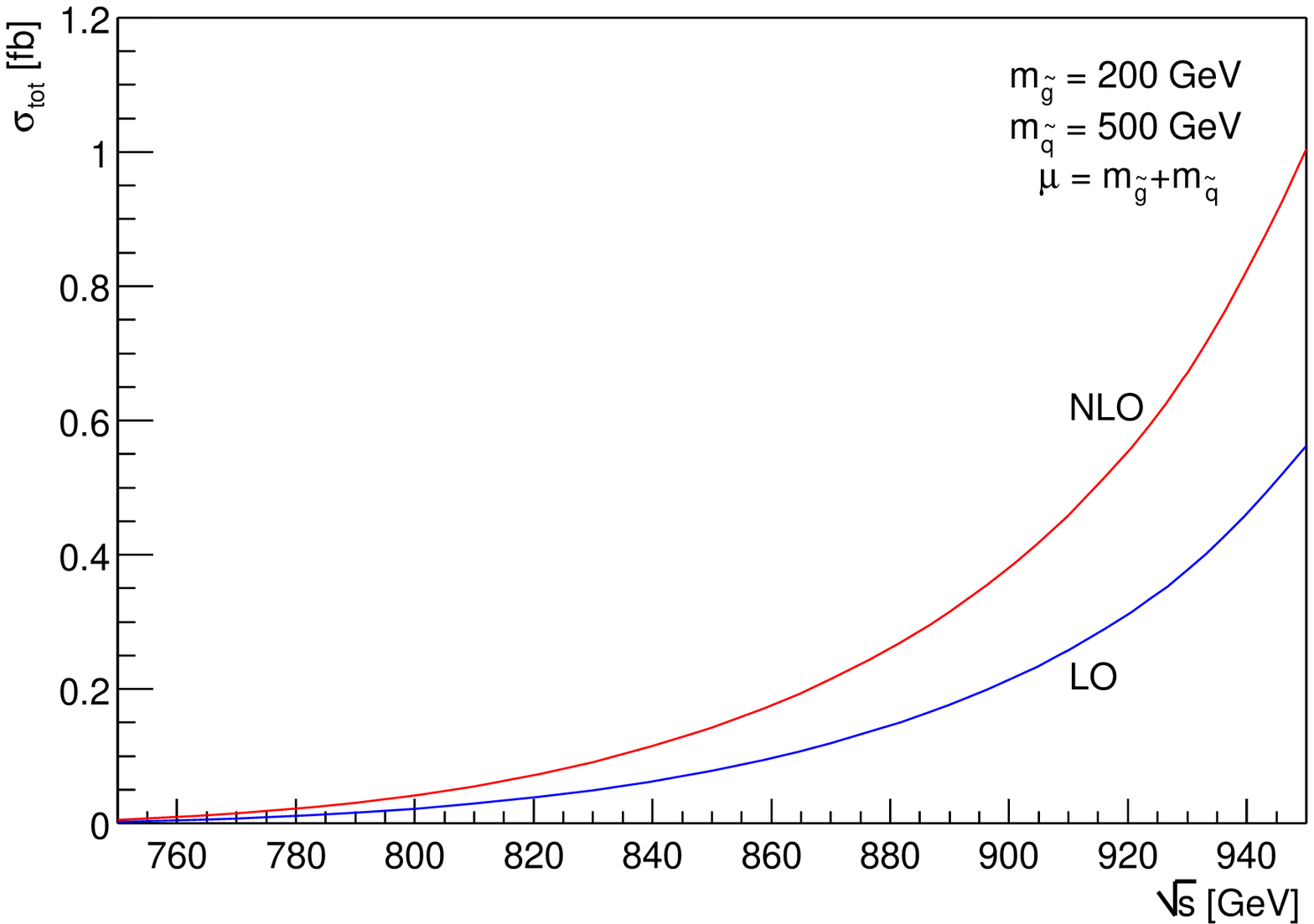,width=8.75cm}}
\put(-2.5,4.75){(a)}
\put(5.75,4.75){(b)}
\end{picture}
\caption{Total cross section for gluino production in LO and NLO for
$m_{\tilde{q}}<m_{\tilde{g}}$ (a) and $m_{\tilde{q}}>m_{\tilde{g}}$ (b).}
\label{marcus1}
\end{center}
\end{figure} 
\begin{figure}[h!]
\unitlength1.0cm
\begin{center}
\vspace{-3cm}
\begin{picture}(8.75,8.75)
\put(-4.,0){\psfig{figure=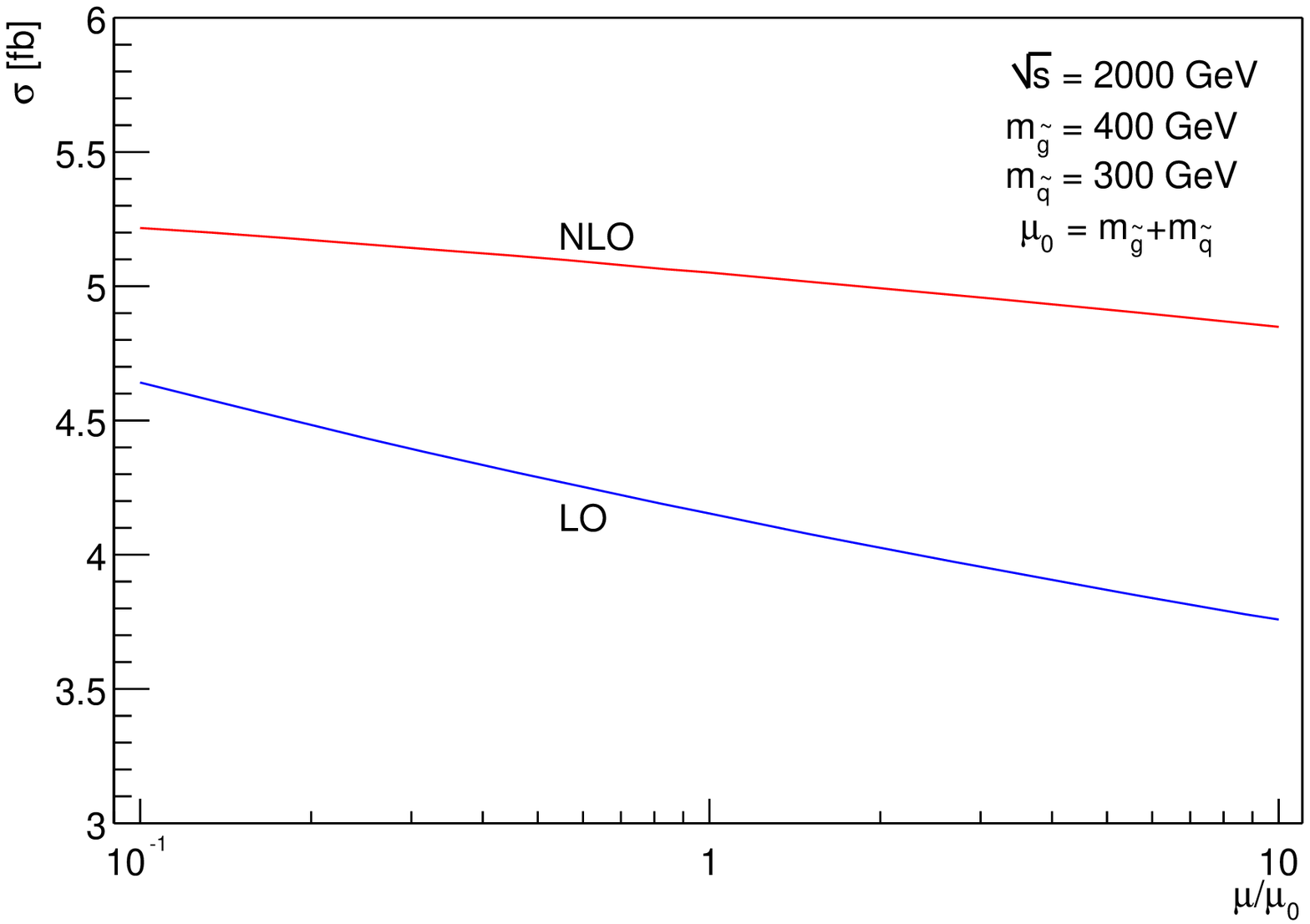,width=8.75cm}}
\put(4.25,0){\psfig{figure=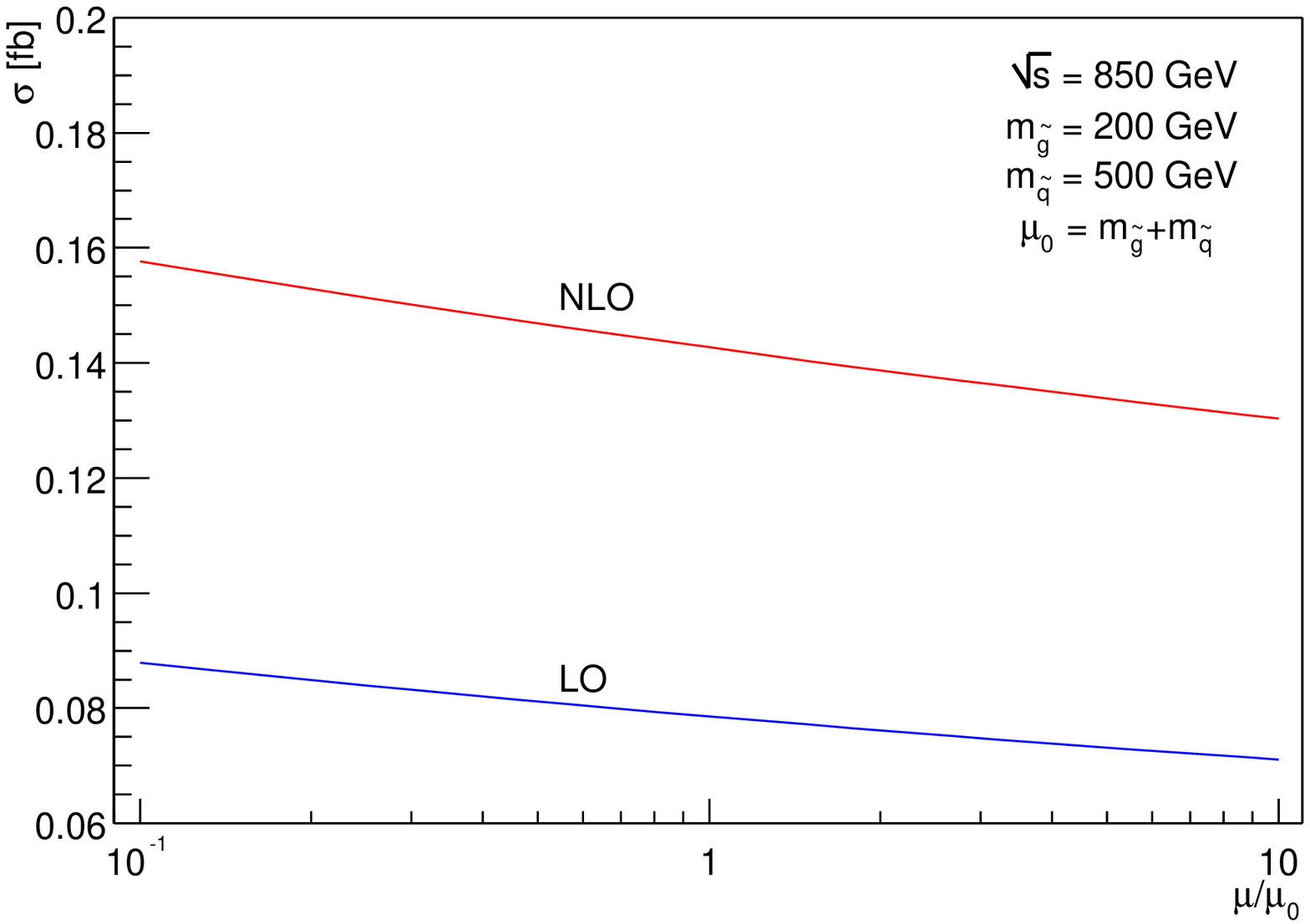,width=8.75cm}}
\put(-2.5,4.75){(a)}
\put(5.75,4.75){(b)}
\end{picture}
\caption{Scale dependence of $\sigma_{\rm tot}$ in LO and NLO
for the cases $m_{\tilde{q}}<m_{\tilde{g}}$ (a) and 
$m_{\tilde{q}}>m_{\tilde{g}}$ (b). The reference scale is 
$\mu_0=m_{\tilde{q}}+m_{\tilde{g}}$ in both cases.}
\label{marcus2}
\end{center}
\end{figure}   
The scale dependence of $\sigma_{\rm tot}$ is shown in Figs.~7a and 7b.
For the phenomenologically more interesting case $m_{\Tilde{q}}<m_{\Tilde{g}}$
the inclusion of the SUSY-QCD corrections significantly reduces the 
theoretical uncertainty due to the arbitrariness 
of the scale choice.
\section{SUSY-QCD corrections to  $e^+e^-\to q\bar{q}g$}
Finally we study the impact of SUSY-QCD corrections
on the production of 3 jets. For this we have to compute the 
SUSY-QCD loop corrections to the process $e^+e^-\to q\Bar{q}g$.
The 3-jet cross section is defined as follows:
\begin{eqnarray}
\sigma_3(y_{\rm cut})=\sigma(e^+e^-\to 3 {\rm\  jets}; y_{ij}>y_{\rm cut}),
\end{eqnarray} 
where we take $y_{ij}=(p_i+p_j)^2/s$ with $p_i$ denoting the four
momenta of the final state partons. The cross section may be expanded
as follows:
\begin{eqnarray}
\sigma_3(y_{\rm cut})=\frac{\alpha_s}{2\pi}\sigma_3^0
+\left(\frac{\alpha_s}{2\pi}\right)^2\left[\sigma_3^{1,\rm SUSY}
+\sigma_3^{1,\rm QCD}\right] +O\left(\alpha_s^3\right).
\end{eqnarray}  
Here $\sigma_3^{1,\rm QCD}$ denotes the ordinary QCD correction
at order $\alpha_s^2$. The  `genuine' SUSY-QCD corrections are 
defined by
\begin{eqnarray}
\Delta_{\rm SUSY}\equiv \frac{\alpha_s}{2\pi}
\frac{\sigma_3^{1,\rm SUSY}}{\sigma_3^0}.
\end{eqnarray} 
The dependence of $\Delta_{\rm SUSY}$ on $y_{\rm cut}$ turns out to be very
small. We choose $y_{\rm cut}=(50{\rm \ GeV})^2/s$ and plot in Fig.~8 
$\Delta_{\rm SUSY}$ as a function of $\sqrt{s}$ for 
$m_{\Tilde{q}}=300$ GeV and $m_{\Tilde{g}}=400$ GeV. 
Here we have used a 
decoupling scheme for the strong coupling, i.e. only the light quark
flavours contribute to the running of $\alpha_s$. This ensures that
$\Delta_{\rm SUSY}\to 0$ as the squark and gluino masses become
large compared to the c.m. energy. Fig.~8 shows that the virtual SUSY-QCD
corrections are tiny at the $Z$-pole, reach a maximum of about 1.5\% around
$\sqrt{s}=1$ TeV and eventually turn negative for large c.m. energies. 
 \begin{figure}
\begin{center}
\vspace{-0.5cm}
\psfrag{yaxislabel}{$\Delta_{\rm SUSY}$}
\epsfig{file=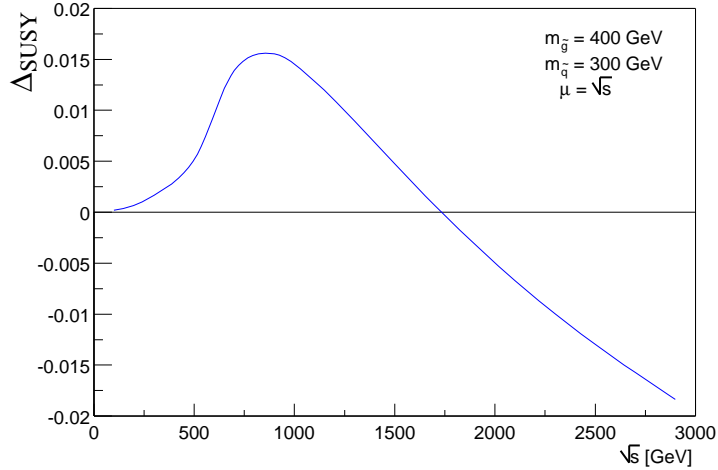,width=10cm}
\caption{Impact of virtual SUSY-QCD corrections
on the 3-jet cross section. We require $(p_i+p_j)^2> (50\ {\rm GeV})^2$ for
each pair $(i,j)$ of the final state partons.}
\label{qqg}
\end{center}
\end{figure}
\section{Conclusions}
The experimental verification of supersymmetric 
coupling relations is necessary to establish supersymmetry.
The SU(3) coupling relation can be tested in $e^+e^-$ collisions
by comparing measurements of the processes $e^+e^-\to
q\Bar{q}g,\ \Tilde{q}\Bar{\Tilde{q}}g,\ 
q\Bar{\Tilde{q}}\Tilde{g}/\Bar{q}\Tilde{q}\Tilde{g}$
to precise, i.e. NLO calculations. 
We have computed the previously unknown order $\alpha_s^2$ SUSY-QCD 
corrections to the above processes. The corrections enhance the
LO cross sections for $e^+e^-\to \Tilde{q}\Bar{\Tilde{q}}g,\ 
q\Bar{\Tilde{q}}\Tilde{g}/\Bar{q}\Tilde{q}\Tilde{g}$ by about 20\% or more in the peak region. The renormalization scale
dependence is reduced for the cases of phenomenological interest. 
\section*{Acknowledgments}
We thank P. M. Zerwas for his collaboration on this project. 

\end{fmffile}
\end{document}